\documentclass[11pt,graphicx,amsmath]{article}
\usepackage{amsmath}
\usepackage{graphicx}
\usepackage{bm}
\usepackage[dvips]{color}

\title{Observational constraints on a Yang-Mills condensate dark energy model}

\author{\small   Z.W. Fu$^1$\thanks{fuzhao@mail.ustc.edu.cn}\ ,
              Y. Zhang$^1$\thanks{yzh@ustc.edu.cn}\ ,
             M.L. Tong$^{2,}$$^1$\thanks{mltong@mail.ustc.edu.cn}\\
       \small $^1$Key Laboratory for Researches in Galaxies and Cosmology, \\
     \small Department of  Astronomy,  University of Science and Technology of China, \\
     \small Hefei, Anhui, 230026,  China \\
     \small $^2$Korea Astronomy and Space Science Institute, Daejeon 305-348, Korea}
 \date{}

\begin{document}
\maketitle

\def\be{\begin{equation}}
\def\ee{\end{equation}}
\def\ba{\begin{eqnarray}}
\def\ea{\end{eqnarray}}
\def\nn{\nonumber}

 \baselineskip=19truept

 \Large

\begin{center}
\Large  Abstract
\end{center}

\begin{quote}

{
Using the recently released Union2 compilation
with 557 Type Ia supernovae,
the shift parameter of cosmic microwave background
given by the WMAP7 observations,
and the baryon acoustic oscillation  measurement from
the Sloan Digital Sky Survey,
we perform the $\chi^2$ analysis on the 1-loop Yang-Mills condensate (YMC)
dark energy model.
The analysis has been made for
 both non-coupling and coupling models
with  $\Omega_{m0}$ and $w_0$ being treated as
free parameters.
It is found that,  $\chi^2_{min}$ = 542.870
at $\Omega_{m0}$ = 0.2701 and $w_0$ = -0.9945 for non-coupling model,
and $\chi^2_{min}$ = 542.790 at
 $\gamma$ = -0.015, $\Omega_{m0}$ = 0.2715 and $w_0$ = -0.9969
  for coupling model.
Comparing with the $\Lambda$CDM model,
the YMC  model has a smaller $\chi^2_{min}$,
but it has greater values of
the  Bayesian and Akaike information criteria.
Overall, YMC is as robust as $\Lambda$CDM.
  }
\end{quote}


\newpage

\section{ Introduction}

The discovery of the current accelerating expansion of the Universe
via the observations of the SNIa \cite{Riess} with follow-ups \cite{Tonry},
further supported by the observations on CMB  \cite{Bennett,WMAP5}
and on large scale structure (LSS) \cite{Bahcall},
has brought a most challenging problem
on both cosmology and physics.
An abundance of literature has been devoted to
the issue.
To interpret the cosmic acceleration,
one may either go beyond the general relativity (GR)
and seek an alternative,
such as the effective gravity \cite{raval},
or stay within the framework of GR
and simply attribute the acceleration to
 some mysterious dark energy (DE) as the driving source.
A positive cosmological constant $\Lambda$ is the simplest candidate
of DE, but it is plagued with the coincidence problem
\cite{Weinberg0}. Various dynamic models have been proposed address
this issue \cite{copeland}. One class of models is based upon some
scalar field, such as quintessence \cite{quint}, k-essence \cite{k},
phantom \cite{phantom}, quintom \cite{quintom}, etc.
In our previous works
\cite{zhang0,oneloopint,XiaTong2loop,swang}, we have developed a
vector field type of dynamic DE model, in which the
renormalization-group improved effective
 Yang-Mills Condensate (YMC)  \cite{Adler} serves as
the dynamical DE. This has bee highly motivated by the great success
of Yang-Mills fields as a corner-stone of particle physics, which
mediate interactions between fundamental particles and define the
vacuum structure.
It has been demonstrated that
YMC DE model has the following properties desired for a dynamical model:
being able to solve the coincidence problem naturally,
giving an equation of state (EoS) $w$ that cross
$-1$ smoothly in the coupling case,
having  the dynamic stability,
and alleviating the high-redshift cosmic age problem.
Moreover, these properties are retained by YMC models even
with the increase of order of quantum corrections, i.e, all 1-loop
\cite{oneloopint}, 2-loop \cite{XiaTong2loop}, and 3-loop
\cite{swang} models have exhibited the same dynamical behavior.
For a detailed description of the 1-loop YMC dark energy model,
including its theoretical basis and the dynamics,
see Ref.\cite{oneloopint}.
Other models using spin-1 fields as a candidate for the dark energy
can be found in Refs.\cite{Elizalde}.

Any DE model has to be confronted with observations.
Recently, there are some significant updates in the observations of
SNIa and CMB.
The Union SNe Ia compilation \cite{kowalski}
enlarged by the CfA3 sample \cite{hicken}
has been recently updated, including a few SNe Ia of high redshifts,
and contains 557 SNe Ia, forming the Union2 compilation \cite{Amanullah}.
Not only the number of SNe is substantially increased,
but also the range of redsifts is extended,
crucial for determining the evolution of dynamical dark energy.
Besides,
the 7-year data of Wilkinson Microwave Anisotropy Probe (WMAP)
has given an improved determination of
cosmological parameters  \cite{wmap7},
in combination with the latest distance measurements  \cite{SDSS percival}
of the Baryon  Acoustic  Oscillations (BAO) \cite{Eisenstein}
in the distribution of galaxies of SDDS and  2dFGRS.
Motivated by these, it is natural for us to expand our work \cite{swang}
to further constrain the YMC  model
with these observational datasets.
In this work, we shall carry out a statistical analysis,
utilizing the combination of SNIa, CMB, and LSS data,
to constrain 1-loop YMC model for both non-coupling and coupling cases.
The examination is much more refined
 than the previous work \cite{swang} in performing statistics.
Furthermore, employing
the Bayesian information criterion(BIC) \cite{Schwarz}
 as well as  the Akaike Information criterion(AIC) \cite{Akaike},
we shall also perform extra statistical examinations
for models with different number of parameters.
The resulting statistics shows that,
 comparing with the $\Lambda$CDM model,
the YMC model is still a robust dynamical dark energy model.
The unit system with $c = \hbar = 1$ is used in this paper.

\section{The 1-loop YMC model}

We consider a
spatially flat  ($k=0$) Robertson-Walker (RW) universe,
whose  expansion is determined by the Friedmann equations
\be \label{friedmann1}
H^ 2= \frac{8 \pi G}{3}(\rho_y+\rho_m),
\ee
\be \label{friedmann2}
\frac{\ddot{a}}{a} =
       -\frac{4 \pi G}{3}(\rho_y+3p_y+\rho_m),
\ee
where $H=\frac{\dot{a}}{a}$, $\rho_y$ and $\rho_m$ are
the energy density of the YMC
and the matter (including both baryons and dark matter),
respectively,
$p_y$ is the pressure of the YMC.
For simplicity,
the radiation component is neglected since
its contribution is very small in the matter dominated era under consideration.
 The energy density
$\rho_y$ and the pressure $p_y$ of
the 1-loop YMC are given by \cite{Zhang94,zhang0}
\be \label{rhoyy}
\rho_y = \frac{1}{2}b \kappa^2 (y+1)e^y,
\ee
\be
p_y = \frac{1}{2}b \kappa^2(\frac{1}{3}y-1)e^y,
\ee
where $\kappa$ is the
renormalization scale of dimension of squared mass,
$b=\frac{22}{3(4\pi)^2}$ for the gauge group $SU(2)$ without fermions,
$y\equiv \ln|E^2 / \kappa^2|$ with
$E^2$ being the squared electric field of YM condensate \cite{Adler}.
When one requires that $\rho_y$ be the dynamical dark energy
and its value at  $z=0$ be equal to  $\sim 0.73\rho_c$,
one finds $\kappa^{1/2} \simeq 5\times 10^{-3}$ eV \cite{oneloopint}.
The equation of state (EoS) for the YMC  is
\be  \label{eos}
w =\frac{p_y}{\rho_y} =  \frac{y-3}   {3y+3}.
\ee
The dynamical evolutions of the YMC dark energy and
the matter are given by
\be \label{ymeq1}
\dot{\rho}_y+3\frac{\dot{a}}{a}(\rho_y+p_y)=-\Gamma\rho_y,
\ee
\be \label{meq1}
\dot{\rho}_m+3\frac{\dot{a}}{a}\rho_m=\Gamma\rho_y,
\ee
where $\Gamma$ is a model parameter,
representing phenomenologically the coupling between
the YMC and the matter components.
For $\Gamma>0$,
the interaction term $\Gamma\rho_y$
is the transfer rate of the YMC energy into matter,
whereas $\Gamma<0$ means that the matter
 transfers energy into the YMC.
For computing  convenience,
Eqs.(\ref {friedmann1}), (\ref {ymeq1}) and (\ref{meq1})
can be recast into:
\be \label{hubble}
(\frac{\dot{a}}{a})^2=\frac{4 \pi Gb\kappa^2}{3}h^2,
 \ee
\be \label{xx}
\frac{dx}{d N}=\gamma\frac{(1+y)e^y}{h}-3x,
\ee
\be \label{yy}
\frac{dy}{d N}=-\gamma\frac{1+y}{(2+y) h}-\frac{4y}{2+y},
\ee
where
$x \equiv \rho_m/ \frac{1}{2}b \kappa^2$,
 $N\equiv\ln a(t)$,
 $h\equiv\sqrt{(1+y)e^y+x}$,
and $\gamma\equiv \Gamma/ \sqrt{4 \pi Gb\kappa^2/3}$.

The set of dynamic equations is completely determined by
the initial values $x_i$ and $y_i$ for the noncoupling case,
plus  $\gamma$ in the coupling case.
In actual computation,
the initial condition can be taken at
 a redshift $z_{i}=1092$.
For a given $\gamma $,
each set of  ($x_i$, $y_i$)
yields a corresponding set ($x_0$, $y_0$) at $z=0$
as  the outcome from solving the dynamic equations,
and hence EoS $w_0 =  (y_0-3)/(3y_0+3)$
and the matter fraction
$\Omega_{m0}=x_0/(x_0+(y_0+1)e^{y_0})$.
This point of our treatment differs from
those models like XCDM \cite{haowei},
where $w_X$ or $\Omega_{m0}$ in some model, was put in by hand
as parameters,
instead of following from dynamic equations.
It should also be mentioned that
we will not do statistical examination
with the Hubble parameter  $H_0= 100$ h km sec$^{-1}$Mpc$^{-1}$ in this paper,
($x_0$, $y_0$) can be taken as two independently adjusted parameters
for the statistical examination.

\section{Constraints from SNIa, BAO and CMB}
Below, we confront the YMC model
with the latest observational distance modulus $\mu_{obs}(z_i)$ data
of 557 SNIa \cite{Amanullah},
the BAO measurement from the Sloan Digital Sky Survey (SDSS) \cite{SDSS percival}
and the shift parameter of
CMB updated by the 7-year WMAP observations \cite{wmap7}.
The theoretical distance modulus is defined as
\be
\mu_{th}(z)\equiv5\log_{10}D_L(z)+\mu_0
\ee
 where
\be\mu_0\equiv42.38-5\log_{10}h,
\ee
 and
 \be D_L(z)=(1+z)\int_0^z \frac{dz'}{E(z')}\ee
is  the luminosity distance in a spatially flat Universe,
actually independent of the Hubble constant $H_0$.
For our model,
\be E(z)
=\sqrt{\Omega_{m0}\frac{\rho_m(z)}{\rho_m(0)}
+(1-\Omega_{m0})\frac{\rho_y(z)}{\rho_y(0)}},
\ee
which depends upon the adjusted parameters ($\Omega_{m0}$ , $w_0$,  $\gamma$).
Here $E(z)$ is an implicit function for YMC
through $\rho_m(z)$ and $\rho_y(z)$
as the solution from the dynamical equations (\ref{hubble})-(\ref{yy}).
In doing the statistical examination in the following,
an ensemble of the solutions of the equations are generated,
each of which correspond  to a point in
the grid of ($\Omega_{m0}$ , $w_0$,  $\gamma$).
This requires much more computing time than the models like XCDM  with
an expression for $E(z)$ containing
parameters explicitly \cite{haowei,Nesseris}.

Note that, to reveal the possible dynamical property of dark energy,
we choose the EoS $w_0$ as free parameter,
which is of dynamical nature and reflects the second order time derivative
$\ddot a(t)$.
This is pertinent since the Union2 dataset
provides SN Ia with higher redshifts,
better for constraining the evolutional property
of a DE model.
For examining a dynamical DE  model,
this has the advantage to the choice of
the Hubble parameter $h$ in Ref.\cite{swang}.
More importantly,
the constraints on cosmological parameters
by the BAO measurement from SDSS and 2dF Galaxy Redshift Survey (2dFGRS)
has an intrinsic degeneracy with ($\Omega_{m0}, h$).
Therefore, one can not arrive at reliable estimate
for both $\Omega_{m0}$ and $h$  simultaneously \cite{SDSS percival,Eisenstein}.
As a matter of fact,
if one would want to determine $h$ with sufficient accuracy,
one has to go beyond
and to employ some extra data,
such as  the HST's key project  \cite{Hubble para}
and SDSS for the history of the Hubble parameter \cite{Gaztanaga}, etc.

For the  SNIa data,
the corresponding $\chi^2$ estimator is
constructed as:
\be\label{chi2SN}
\chi^2_{SN}({\bf p};\mu_0)
       =\sum_{i=1}^{557}\frac{[\mu_{obs}(z_i)-\mu_{th}(z_i)]^2}{\sigma_i^2},
\ee
where $\bf p$ stands for a set of parameters,
such as ($\Omega_{m0}$, $w_0$,  $\gamma$),
$\mu_{obs}(z_i)$ is the observed value
of distance modulus for the $i$th supernova,
and $\sigma_i $ is the corresponding $1\sigma$ error that
can  be found via Ref.\cite{Amanullah}.
The nuisance parameter $\mu_0$ can be analytically marginalized
over \cite{Nesseris},
so that one actually minimizes $\chi^2_{SN}(\bf{p})$
instead of  $\chi^2_{SN}({\bf p};\mu_0)$.
The minimization with respect to $\mu_0$
can be made simply by expanding the $\chi^2$ of Eq.
(\ref{chi2SN}) with respect to $\mu_0$ as
\begin{equation}\label{chi2}
\chi^2_{SN}({\bf p})= A({\bf p})-2\mu_0B({\bf p})+\mu_0^2 C,
\end{equation}
where
\begin{equation}
A({\bf p})=\sum\limits_{i=1}^{557}{[\mu_{obs}(z_i)-\mu_{th}(z_i;\mu_0=0,{\bf p})]^2\over \sigma_i^2},
\end{equation}
\begin{equation}
B({\bf p})=\sum\limits_{i=1}^{557}{\mu_{obs}(z_i)-\mu_{th}(z_i;\mu_0=0,{\bf p})\over \sigma_i^2},
\end{equation}
\begin{equation}
C=\sum\limits_{i=1}^{557}{1\over \sigma_i^2}.
\end{equation}
Evidently, Eq.(\ref{chi2}), as well as (\ref{chi2SN}),
 has a minimum for $\mu_0=B/C$ at
\begin{equation}
\tilde{\chi}^2_{SN}({\bf p})=A({\bf p})-{B({\bf p})^2\over C}.\label{tchi2sn}
\end{equation}
Since $\chi^2_{SN, min} = \tilde{\chi}^2_{SN, min}$, instead of
minimizing $\chi^2_{SN}$ one can minimize $\tilde{\chi}^2_{SN}$,
which is now independent of the nuisance parameter $\mu_0$.
Notice that this analytical marginalization over $\mu_0$
means that the Hubble parameter $h$ is effectively marginalized over.

Next, the distance parameter $A$ of the measurement of
 the BAO peak in the distribution of SDSS luminous red
 galaxies is defined as  \cite{Eisenstein}
\be
A\equiv \Omega_m^{1/2}E(z_b)^{-1/3}
      \left[\frac{1}{z_b}\int_0^{z_b}\frac{dz}{E(z)}\right]^{2/3}
\ee
 with $z_b=0.35$,
 the redshift at which the acoustic scale has been measured.
The observations \cite{Eisenstein}  give
 \be A_{obs} = 0.469(n_s/0.98)^{-0.35}\pm0.017,
  \ee
 where $n_s $  is the primordial spectral index and
WMAP7 data \cite{wmap7} yields  the updated value $n_s= 0.963$,
while it was $n_s= 0.960$ by WMAP5 \cite{WMAP5}.
The corresponding $\chi^2$ of BAO is given by
\be
\chi_{BAO}^2 = \frac{(A-A_{obs})^2}{\sigma_A^2},
\ee
where $\sigma_A = 0.017$.

Finally,
the shift parameter $R$ from CMB
is defined as
 \be R\equiv\Omega_{m0}^{1/2}\int_0^{z_{rec}}\frac{dz}{E(z)}
\ee
with $z_{rec}$ being the redshift of recombination.
WMAP7 \cite{wmap7}  gives $ z_{rec}= 1091.3\pm 0.91$ and $ R_{obs} = 1.725 \pm0.018$,
while it was $ z_{rec}= 1090.0\pm 0.93$ and $ R_{obs} = 1.710 \pm 0.019$ by WMAP5 \cite{WMAP5}.
The $\chi^2$ of CMB is
\be
\chi_{CMB}^2 = \frac{(R-R_{obs})^2}{\sigma_R^2},
\ee
with $\sigma_R=0.018$.
As usual, assuming these three sets data of observations
are mutual independent,
and the measurement errors for each set are Gaussian
with the likelihood function of the form
\begin{equation}\label{likelihood}
{\cal{L } } \propto e^{-\chi^2/2}.
\end{equation}
The three sets data are combined by multiplying the likelihoods,
and  the combined $\chi^2$  is given by
\be
\chi^2 = \tilde{\chi}^2_{SN}+\chi_{BAO}^2+\chi_{CMB}^2.
\ee
Variations of values of the parameters  ($\Omega_{m0}$, $w_0$,  $\gamma$)
yield respective values of $\chi^{2}$.
A minimal $\chi^{2}$ has been found,
corresponding to a maximal ${\cal{L } }$,
which would be favored by the observations.

\section{Results}
For the non-coupling  $\gamma = 0$ case, there are two parameters
$\Omega_{m0}$ and $w_0$.
In our present computation
of  the confidence level of $\Omega_{m0}$,
the procedure goes as:
First we compute $\chi^2(\Omega_{m0},w_0) $ and the Likelihood
$\cal L = \cal L$ $(\Omega_{m0},w_0) $
as functions of both $\Omega_{m0}$ and $w_0$,
from which, by a standard searching procedure,
follows  the resulting minimum
$\chi^2_{\rm min} = 542.870$ at $\Omega_{m0} = 0.2701$
and $w_0 = -0.9945$.
Next we integrate $ \cal L$ $(\Omega_{m0},w_0) $
 over $w_0$ and derive the Likelihood
$\cal L$ $(\Omega_{m0}) $ as a function of only $\Omega_{m0}$.
The prior cut off  $w_0 = (-0.9999, -0.90115)$ is assumed here.
Then we compute the confidence level of $\Omega_{m0}$
based upon the function  $\cal L$ $(\Omega_{m0}) $.
The result is that $\Omega_{m0} =
0.2701^{+0.0153}_{-0.0121}$ at  $68.3\%$ CL(Confidence Level),
$\Omega_{m0} = 0.2701^{+0.0297}_{-0.0250}$ at  $95.4\%$ CL.
 Since the non-coupling YMC model always has $w_0\ge -1$ \cite{oneloopint},
it is found that the likelihood function ${\cal{L} } (w_0) $
is mainly distributed in a narrow range close to $w_0=-1$,
so that the computation of CL can only be done for an upper
bound of $w_0$.
The result is that $w_0<-0.9773$ at  $68.3\%$ CL,
and $w_0<-0.9579$ at  $95.4\%$ CL.
 The prior cut off  $\Omega_{m0} = (0.2001, 0.3501)$ is assumed here.
Figure 1 shows the details of the C.L. curves in the $(\Omega_{m0}, w_0)$ plane.
We remark that,
in obtaining the values of C.L. for a $\Omega_{m0}$ (or $w_0$)
at the maximum of ${\cal{L } }$,
the integration over $w_0$ (or $\Omega_{m0}$, respectively) has been carried out.

To make a comparison with the $\Lambda$CDM model as a sort of referring model,
we have also done the computing and fitting
of  the $\Lambda$CDM model to the same observational data as above,
yielding  $\chi^2_{min}$ = 542.919 at $\Omega_{m0} = 0.2701$.
The confidence levels are the following:
 $\Omega_{m0} = 0.2701^{+0.0140}_{-0.0135}$ at  $68.3\%$ CL,
$\Omega_{m0} = 0.2701^{+0.0283}_{-0.0264}$ at  $95.4\%$ CL.
We present the corresponding $\chi^2$ and 
likelihood ${\cal{L } } \propto e^{-\chi^2/2}$ in figure 3.

Based on the $\chi^2$ estimator only, one would draw a conclusion that
the non-coupling YMC model
is more favored in confronting this set of observations,
as it has a slightly smaller $\chi^2_{min}$ than the $\Lambda$CDM model.
A similar result was obtained for non-coupling YMC model at 3-loop
for an earlier dataset of the 182 SNIa +CMB+BAO \cite{swang}.
This kind of better performance of YMC
is due to its dynamical nature,
say, its $w(z)$ being a function of $z$,
instead of a constant $-1$ as in $\Lambda$CDM.
In this regards,
more observational data at high redshifts are desired
to distinguish various DE models
with different evolutionary behavior,
by using certain diagnosis, such as the Statefinder and $Om$  \cite{Sahni}.

\begin{figure}[h!]
\centerline{
\includegraphics[width=8cm]{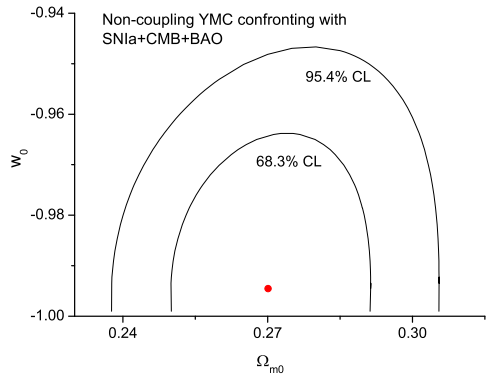}}
 \caption{\label{fig1}
  Non-coupling YMC:  the C.L. curves in $\Omega_{m0}-w_0$ plane
  based upon the joint study of SNIa, BAO and CMB.  }
\end{figure}

\begin{figure}[h!]
\centerline{
 \includegraphics[width=8cm]{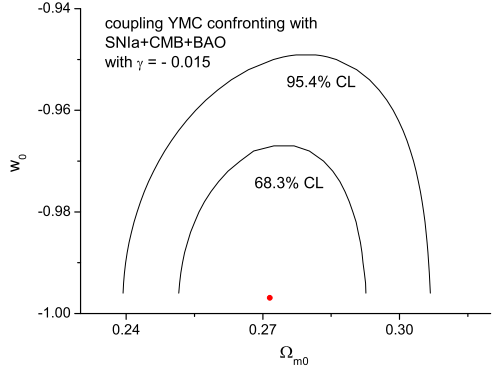}}
 \caption{\label{fig2}
  Coupling YMC  with $\gamma=-0.015$: the C.L. curves in $\Omega_{m0}-w_0$ plane
 based upon the joint of SNIa, BAO and CMB.  }
\end{figure}

\begin{figure}[h!]
\centerline{\includegraphics[width=8cm]{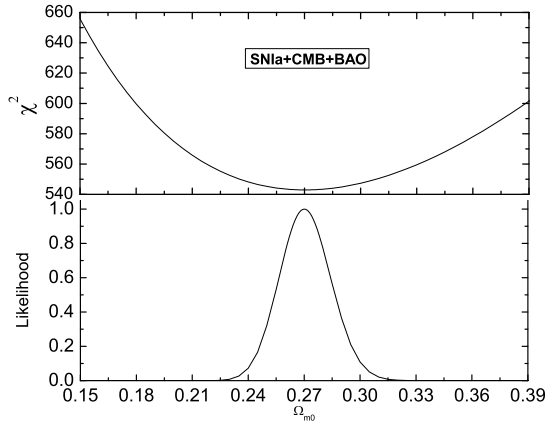}}
 \caption{\label{fig3}
 The $\chi^2$ and $\cal{L}$ of $\Lambda$CDM
 based upon the joint of SNIa, BAO and CMB}
\end{figure}

For the coupling model,
there are three independently adjusted parameters
$\gamma$, $\Omega_{m0}$ and $w_0$.
In order to search for possible DE models in an extended domain,
in particular,
we will allow the coupling $\gamma$ to take negative values.
For each fixed value of $\gamma$, we minimize $\chi^2$
with $\Omega_{m0}$ and $w_0$.
We have done ten different values of $\gamma$,
and the resulting values of $\chi^2_{min}$ for
the corresponding sets of
($\gamma$, $\Omega_{m0}$  $w_0$) are listed in Table 1.
It is seen that, as a function of $\gamma$,
the minimum $\chi^2_{min}=542.790$ is
 attained at $\gamma=-0.015$,
 $\Omega_{m0}=0.2715$ and $w_0=-0.9969$.
This $\chi^2_{min}$ is slightly smaller than that of
the non-coupling and of the $\Lambda$CDM models.
Table 1 also tells  that,
although the $\chi^2_{min}$ depends on $\gamma$,
the dependence is not very strong.

Our previous work \cite{swang}
did not search for the minimal $\chi^2$
with respect to the parameter $\gamma$
since it calculated only three different values of $\gamma$.

Figure 2 shows the C.L. curves with $\gamma=-0.015$
 in the $(\Omega_{m0}, w_0)$ plane.

\begin{table}

\begin{center}
\begin{tabular}{|l|c|c|c|c|}
\hline\hline
$\chi^2_{min}$ & $\gamma$& $\Omega_{m0}$ &$w_0$\\
\hline
543.419 & -0.05 &0.2755  & -0.9898\\
543.119 & -0.04 & 0.2743  & -0.9918\\
542.912 & -0.03 & 0.2731  & -0.9938\\
542.804 & -0.02 & 0.2720  & -0.9959\\
542.790 & -0.015& 0.2715  & -0.9969\\
542.804 & -0.01 & 0.2710  & -0.9979\\
542.870 & 0     & 0.2701  & -0.9945\\
543.202 & 0.05  & 0.2669  & -0.9797\\
543.496 & 0.1   & 0.2651  & -0.9692\\
544.511 & 0.5   & 0.2636  & -0.9313\\
\hline
\end{tabular}
\end{center}
\caption{\label{table1}
For each given coupling  $\gamma$,
the values of $\chi^2_{min}$ and
the corresponding best-fit parameters $\Omega_{m0}$ and $w_0$
are listed.
}
\end{table}

Thus the YMC model with a negative coupling $\gamma<0$
does a bit better than the latter two models
in confronting the updated observations.
We remark that
this conclusion was partly hinted in the 3-loop YMC model,
where the consideration was confined only to
the limited region of $\gamma>0$  \cite{swang}.
Note that $\gamma<0$ means a situation in which
the matter is  transferring energy into the YMC,
and EoS $w_0$ will not cross over $-1$ \cite{oneloopint}.
Besides,
in the far future when the scale factor is hundreds times the present one,
the matter density $\rho_m$ will turn into negative,
which would be a non-physical region.
Thus, the  model of a negative coupling
is phenomenological and can not be infinitely extended into far future.
In regards to the issue of coupling,
we notice that
a positive interaction is preferred
in a study of dynamics of galaxy clusters
with a coupling between dark energy and dark matter \cite{abdalla}.

The $\chi^2$ analysis is effective in searching for
the best-fit values of parameters within a given model.
But for models with different number of parameters,
one would expect $\chi^2_{min}$ decreases
as the number of free model parameters increase.
For model comparisons in this case,
one can use other kinds of criteria for model selection,
such as BIC \cite{Schwarz},  AIC \cite{Akaike}, and
$\chi^2_{min}/dof$ with the degree of freedom $dof = N-k$,
whereas $N$ and $k$ are
the number of data points and the dimension
(number of independently adjusted parameters) of the statistical model, respectively.
The BIC is defined as
$BIC \equiv-2\ln{\cal L}_{max}+k\ln N\,$,
where ${\cal L}_{max}$ is the maximum likelihood.
In the Gaussian case, $\chi^2_{min} = -2\ln{\cal L}_{max}+constant$,
so that the difference in BIC is given by
 $\Delta{\rm BIC}  = \Delta\chi^2_{min}+\Delta k \ln N$.
The AIC is defined as   $AIC \equiv-2\ln{\cal L}_{max}+2k\,$,
 and the difference in AIC is given by
 $\Delta{\rm AIC}=\Delta\chi^2_{min}+2\Delta k$.
We present the results for these three criteria in Table 2,
which imply that
the $\Lambda CDM$ is still favored due to its simplicity,
whereas the coupling model is ranked last as it has most parameters.
\begin{table}
\begin{center}
\begin{tabular}{|l|c|c|c|c|}
\hline\hline
Model & $\Lambda $CDM & non-coupling & coupling\\
\hline
$\chi^2_{min}$  & 542.919 & 542.870 & 542.790\\
$k$   & 1       & 2       & 3\\
$\chi^2_{min}/dof$ & 0.973 & 0.975  & 0.976\\
$\triangle BIC$& 0 & 6.277  & 12.523\\
$\triangle AIC$ & 0 & 1.951  & 3.871\\
Rank     & 1 & 2 & 3\\
\hline
\end{tabular}
\end{center}
\caption{\label{table2} comparison of models  }
\end{table}

In summary, using  the $\chi^2$ analysis alone in confronting to the
updated observational data of  557 SNIa +CMB+BAO, we find that the
non-coupling YMC is favored over $\Lambda$CDM, and the coupling YMC
with a negative $\gamma$ is the most favored.
Taking into
account of the dimension of a statistical model,
$\Lambda$CDM is still simplest.
Overall, YMC is as robust as $\Lambda$CDM.
More observational data at higher redshifts
in future are much desired to shed light on the
issue of dark energy.

\
\

{ACKNOWLEDGMENT}: Y. Zhang's research work has been supported by the
CNSF No.11073018, SRFDP, and CAS.
 We would like to thank W. Zhao for
many helpful discussions.

\end{document}